\documentclass[aps,twocolumn,showpacs]{revtex4}

\newcommand{\stac}[2]{\stackrel{\scriptscriptstyle {#1}}{#2}}

\begin{document}

\title{Randall-Sundrum two D-brane model}

\author{Tetsuya Shiromizu$^{(1,2,3)}$, Yoshiaki Himemoto$^{(2)}$ and Keitaro Takahashi$^{(2)}$}

\affiliation{$^{(1)}$Department of Physics, Tokyo Institute of Technology, 
Tokyo 152-8551, Japan}

\affiliation{$^{(2)}$Department of Physics, The University of Tokyo,  Tokyo
113-0033, Japan}

\affiliation{$^{(3)}$Advanced Research Institute for Science and Engineering, 
Waseda University, Tokyo 169-8555, Japan}

\date{\today}

\begin{abstract}
In Randall-Sundrum two D-brane system we derive the gravitational theory on the branes. 
It is turned out from the consistency that one D-brane has the negative tension brane under 
Randall-Sundrum tuning and both gauge fields on the brane are related by scale transformation 
through the bulk RR/NS-NS fields. As same with the single D-brane case, the gauge field which 
is supposed to be localised on the brane does not couple to the gravity on the branes.
\end{abstract}


\maketitle

\label{sec:intro}
\section{Introduction}

Superstring theory provides us new picture of our universe in higher dimensional spacetimes: 
our universe may be a thin domain wall in higher dimensions. 
The simplest models were proposed by Randall and Sundrum \cite{RSI,RSII}. Therein 
the bulk spacetime and the brane are governed by the five dimensional Einstein gravity and 
the Nambu-Goto action, respectively. Although such pictures rely on the feature of D-brane, 
realistic models based on D-brane has not been seriously considered so far. 
Recently a single D-brane model which is the D-brane version of Randall-Sundrum II model \cite{RSII} 
was addressed by long wave approximation in Ref. \cite{SKOT,OSKH}.
Surprisingly, the gauge field which is supposed to localise on the brane cannot be source
for the gravity on the brane when the net cosmological constant on the brane vanishes. 
If the net cosmological constant exists, the gauge fields can be
source for the gravity \cite{SKT}. However, there were undetermined terms which will be fixed by 
the boundary/initial condition near the Cauchy horizons in anti-deSitter spacetime. 
In general there is no guarantee that we can ignore such undetermined terms. Hence we 
will consider two D-brane system in this brief report. In this case the undetermined terms 
can be completely fixed by two junction conditions on the branes. 

The rest of this paper is organised as follows. In Sec. II, we describe the setup of model, 
field equations and junction conditions. In Sec. III, we solve the bulk spacetime using 
long wave approximation \cite{GE} and derive the field equations on the brane.
Finally we give summary and discussion in Sec. IV.

\section{Model}
\label{sec:model}
\subsection{The action for toy model}

For simplicity we work with a toy model. It is a mimic of a theory derived from ten-dimensional 
type IIB supergravity through dimensional reduction to five.
The main differences between the original one \cite{SKOT,ST} and the current toy model 
are as follows. In the current model, there are no scalar fields corresponding 
to dilaton and radius of compactification on $S^5$. 
Instead, we introduced the bulk cosmological constant $\Lambda$. 
In single D-brane case it is turned out that the contribution from scalar fields are 
not so important when 
one wants to discuss the coupling of the gauge field on the brane to the gravity. 
In fact, we will be able to see the same result, that is, the gauge field localised on 
the brane does not couple to the gravity. The cancellation between the 
contribution from the NS-NS and RR fields is essential. See Ref. \cite{SKOT,ST} for the 
original five dimensional action. 

Now we are thinking of two D-brane model and its total action is given by 
%
\begin{eqnarray}
S & = & \frac{1}{2\kappa^2} \int d^5x {\sqrt {-G}}\biggl[{}^{(5)}R-2\Lambda
-\frac{1}{2}|H|^2
\nonumber \\
& & -\frac{1}{2}(\nabla \chi)^2-\frac{1}{2}|\tilde F|^2-\frac{1}{2}|\tilde
G|^2 \biggr] \nonumber \\
& & +S_{\rm brane}^{(+)}+S_{\rm CS}^{(+)}+S_{\rm brane}^{(-)}+S_{\rm CS}^{(-)} ,
\label{action}
\end{eqnarray}
%
where $H_{MNK}=\frac{1}{2}\partial_{[M}B_{NK]}$, 
$F_{MNK}=\frac{1}{2}\partial_{[M}C_{NK]}$, 
$G_{K_1 K_2 K_3 K_4 K_5}=\frac{1}{4!}\partial_{[K_1}D_{K_2 K_3 K_4 K_5]}$, 
$\tilde F = F + \chi H$ and $\tilde G=G+C \wedge H$. $M,N,K=0,1,2,3,4$. 
$B_{MN}$ and $C_{MN}$ are 2-form fields, and $D_{K_1 K_2 K_3 K_4}$ is
a 4-form field. $\chi$ is a scalar field.

$S_{\rm brane}^{(\pm)}$ is given by Born-Infeld action
%
\begin{eqnarray}
S_{\rm brane}^{(+)}=\gamma_{(+)} \int d^4x {\sqrt {-{\rm det}(h+{\cal F}^{(+)})}},
\end{eqnarray}
%
%
\begin{eqnarray}
S_{\rm brane}^{(-)}=\gamma_{(-)} \int d^4x {\sqrt {-{\rm det}(q+{\cal F}^{(-)})}},
\end{eqnarray}
%
where $h_{\mu\nu}$ and $q_{\mu\nu}$ are the induced metric on the $D_{\pm}$-brane and 
%
\begin{eqnarray}
{\cal F}_{\mu\nu}^{(\pm)}=B_{\mu\nu}^{(\pm)}+(-\gamma_{(\pm)})^{-1/2}F_{\mu\nu}^{(\pm)}.
\end{eqnarray}
%
$F_{\mu\nu}$ is the $U(1)$ gauge field on the brane. $\mu,\nu=0,1,2,3$. 
$S_{\rm CS}^{(\pm)}$ is Chern-Simons action 
%
\begin{eqnarray}
S_{\rm CS}^{(+)} & = & \gamma_{(+)} \int d^4x {\sqrt {-h}}
\epsilon^{\mu\nu\rho\sigma}\biggl[ \frac{1}{4}{\cal
F}_{\mu\nu}^{(+)}C_{\rho\sigma}^{(+)}+\frac{\chi}{8}{\cal F}_{\mu\nu}^{(+)}{\cal
F}_{\rho\sigma}^{(+)}
\nonumber \\
& & +\frac{1}{24}D_{\mu\nu\rho\sigma}^{(+)} \biggr],
\end{eqnarray}
%
%
\begin{eqnarray}
S_{\rm CS}^{(-)} & = & \gamma_{(-)} \int d^4x {\sqrt {-q}}
\epsilon^{\mu\nu\rho\sigma}\biggl[ \frac{1}{4}{\cal
F}_{\mu\nu}^{(-)}C_{\rho\sigma}^{(-)}+\frac{\chi}{8}{\cal F}_{\mu\nu}^{(-)}{\cal
F}_{\rho\sigma}^{(-)}
\nonumber \\
& & +\frac{1}{24}D_{\mu\nu\rho\sigma}^{(-)} \biggr].
\end{eqnarray}
%

\subsection{Basic equations}

In this subsection we write down the basic equations and boundary conditions. 
Let us perform (1+4)-decomposition 
%
\begin{eqnarray}
ds^2=G_{AB}dx^{A}dx^{B}=e^{2\phi (x)}dy^2+g_{\mu\nu}(y,x) dx^\mu dx^\nu,
\end{eqnarray}
%
where $y$ is the coordinate orthogonal to the brane and $\mu, \nu = 0,1,2,3$.
$D_+$-brane and $D_-$-brane are supposed to locate at $y=0$ and $y=y_0$. 
For simplicity, we assume $\tilde F_{\mu\nu\alpha} =0$ and $H_{\mu\nu\alpha}=0$, that is, 
they are closed. 
 
The ``evolutional" equations to the $y$-direction are 
%
\begin{eqnarray}
e^{-\phi} \partial_y K 
& = & {}^{(4)} R-\kappa^2 \biggl( {}^{(5)}T^\mu_\mu -\frac{4}{3}{}^{(5)}T^M_M \biggr) -K^2
\nonumber \\
& & -e^{-\phi}D^2 e^\phi, 
\end{eqnarray}
%
%
\begin{eqnarray}
e^{-\phi} \partial_y \tilde K^\mu_\nu & = &  {}^{(4)}\tilde R^\mu_\nu 
-\kappa^2\biggl({}^{(5)}T^\mu_\nu
-\frac{1}{4}
\delta^\mu_\nu {}^{(5)}T^\alpha_\alpha \biggr)-K \tilde K^\mu_\nu \nonumber \\
& & ~~-e^{-\phi}[D^\mu D_\nu e^{\phi}]_{\rm traceless},
\label{traceless}
\end{eqnarray}
%
%
\begin{eqnarray}
\partial_y^2 \chi +D^2 \chi +e^\phi K\partial_y \chi-\frac{1}{2}H_{y\alpha\beta}\tilde 
F^{y\alpha\beta}=0,
\end{eqnarray}
%
%
\begin{eqnarray}
\partial_y X^{y\mu\nu}+e^\phi KX^{y\mu\nu}
+\frac{1}{2}F_{y\alpha\beta}\tilde G^{y\alpha\beta\mu\nu}=0,
\end{eqnarray}
%
%
\begin{eqnarray}
\partial_y \tilde F^{y\mu\nu}+e^\phi K \tilde F^{y\mu\nu}-\frac{1}{2}H_{y\alpha\beta}
\tilde G^{y\alpha\beta\mu\nu}=0,
\end{eqnarray}
%
%
\begin{eqnarray}
\partial_y \tilde G_{y \alpha_1 \alpha_2 \alpha_3 \alpha_4}
=e^\phi K\tilde  G_{y \alpha_1 \alpha_2 \alpha_3 \alpha_4},
\end{eqnarray}
%
where $X^{y\mu\nu}:=H^{y\mu\nu}+\chi \tilde F^{y\mu\nu}$ and the 
energy-momentum tensor is
%
\begin{eqnarray}
&& \kappa^2\;{}^{(5)\!}T_{MN} =  \frac{1}{2}\biggl[ \nabla_M \chi \nabla_N \chi
-\frac{1}{2}g_{MN} (\nabla \chi)^2 \biggr]
\nonumber \\
& & ~~~~~~~~~~
+\frac{1}{4}\biggl[H_{MKL}H_N^{~KL}-g_{MN}|H|^2 \biggr] 
\nonumber \\
& & ~~~~~~~~~~
 +\frac{1}{4}\biggl[\tilde F_{MKL}\tilde
F_N^{~KL}-g_{MN}|\tilde F|^2
\biggr]
\nonumber \\
& & ~~~~~~~~~~
 +\frac{1}{96}\tilde G_{MK_1 K_2 K_3 K_4} \tilde G_{N}^{~~K_1
K_2 K_3 K_4}-\Lambda g_{MN}.
\nonumber \\
& & 
\end{eqnarray}
%
$K_{\mu\nu}$ is the extrinsic curvature, $K_{\mu\nu}=\frac{1}{2}e^{-\phi} \partial_y g_{\mu\nu}$. 
$\tilde K^\mu_\nu$ and ${}^{(4)}\tilde R^\mu_\nu$ are the traceless parts
of $K^\mu_\nu$ and ${}^{(4)}R^\mu_\nu$, respectively.

The constraints are 
%
\begin{eqnarray}
& & -\frac{1}{2}\biggl[{}^{(4)}R-\frac{3}{4}K^2+\tilde K^\mu_\nu \tilde K^\nu_\mu \biggr]
=\kappa^2\:{}^{(5)\!}T_{yy}e^{-2\phi}, 
\end{eqnarray}
%
%
\begin{eqnarray}
D_\nu K^\nu_\mu-D_\mu K = \kappa^2\:{}^{(5)\!}T_{\mu y}e^{-\phi},
\end{eqnarray}
%
%
\begin{eqnarray}
D^\alpha(e^{-\phi} X_{y\alpha\mu})=0, \label{con1}
\end{eqnarray}
%
%
\begin{eqnarray}
D^\alpha (e^{-\phi} \tilde F_{y\alpha\mu})=0, \label{con2}
\end{eqnarray}
%
%
\begin{eqnarray}
D^\alpha (e^{-\phi} \tilde G_{y \alpha \mu_1 \mu_2 \mu_3})=0,
\end{eqnarray}
%
where $D_\mu$ is the covariant derivative with respect to $g_{\mu\nu}$. 

Under $Z_2$-symmetry, the junction conditions at the brane located $y=0$ are 
%
\begin{eqnarray}
& & K_{\mu\nu}-h_{\mu\nu}K  =  
\frac{-\kappa^2 \gamma_{(+)}}{2}(h_{\mu\nu}-T_{\mu\nu}^{(+)}) 
+O(T_{\mu\nu}^2) \\
& & H_{y\mu\nu}(0,x)=-\kappa^2 \gamma_{(+)} e^\phi {\cal F}_{\mu\nu}^{(+)}, \\
& & \tilde F_{y\mu\nu}(0,x)
=-\frac{\kappa^2}{2}\gamma_{(+)} e^\phi \epsilon_{\mu\nu\alpha\beta}{\cal F}^{(+)\alpha\beta}, \\
& & \tilde G_{y\mu\nu\alpha\beta}(0,x)
=- \kappa^2 \gamma_{(+)} e^\phi \epsilon_{\mu\nu\alpha\beta},\\
& & \partial_y \chi (0,x) 
= -\frac{\kappa^2}{8}\gamma_{(+)} e^\phi \epsilon^{\mu\nu\alpha\beta}{\cal F}^{(+)}_{\mu\nu}{\cal
F}_{\alpha\beta}^{(+)}.
\end{eqnarray}
%
The junction conditions at the brane located $y=y_0$ are 
%
\begin{eqnarray}
& & K_{\mu\nu}-q_{\mu\nu}K  =  
\frac{\kappa^2 \gamma_{(-)}}{2}(q_{\mu\nu}-T_{\mu\nu}^{(-)}) +O(T_{\mu\nu}^2) \\
& & 
H_{y\mu\nu}(y_0,x)=\kappa^2 \gamma_{(-)} e^\phi {\cal F}_{\mu\nu}^{(-)}, \\
& & \tilde F_{y\mu\nu}(y_0,x)
=\frac{\kappa^2}{2}\gamma_{(-)} e^\phi \epsilon_{\mu\nu\alpha\beta}{\cal F}^{(-)\alpha\beta},\\
& & \tilde G_{y\mu\nu\alpha\beta}(y_0,x)
=\kappa^2 \gamma_{(-)} e^\phi \epsilon_{\mu\nu\alpha\beta}, \\
& & \partial_y \chi (y_0,x) 
= \frac{\kappa^2}{8}\gamma_{(-)} e^\phi \epsilon^{\mu\nu\alpha\beta}{\cal F}^{(-)}_{\mu\nu}{\cal
F}_{\alpha\beta}^{(-)}.
\end{eqnarray}
%
In the above 
%
\begin{eqnarray}
T^{(\pm)\mu}_{~~~\nu}={\cal F}^{(\pm)\mu\alpha}{\cal F}^{(\pm)}_{\alpha\nu} -\frac{1}{4}\delta^\mu_\nu
{\cal F}_{\alpha\beta}^{(\pm)} {\cal F}^{(\pm) \alpha\beta}.
\end{eqnarray}
%

\section{Long-wave approximation}
\label{sec:approximation}

In this section, we approximately solve the bulk field equations by long wave approximation(gradient 
expansion \cite{GE}). 

The bulk metric is written again as,
%
\begin{eqnarray}
ds^2=e^{2\phi (x)}dy^2+g_{\mu\nu}(y,x) dx^\mu dx^\nu. 
\end{eqnarray}
%
The induced metric on the brane will be denoted by 
$h_{\mu\nu}:=g_{\mu\nu}(0,x)$ 
and then
%
\begin{eqnarray}
g_{\mu\nu}(y,x)
=a^2(y,x)\Bigl[h_{\mu\nu}(x)+\stac{(1)}{g}_{\mu\nu}(y,x)+\cdots\Bigr].
\end{eqnarray}
%
In the above $\stac{(1)}{g}_{\mu\nu}(0,x)=0 $ and $a(0,x)=1$. In a similar way, 
the extrinsic curvature is expanded as 
%
\begin{eqnarray}
K^\mu_\nu = \stac{(0)}{K^\mu_\nu}+ \stac{(1)}{K^\mu_\nu}+\stac{(2)}{K^\mu_\nu}+
\cdots.
\end{eqnarray}
%
The small parameter is $\epsilon = (\ell /L)^2 \ll 1$,
where $L$ and $\ell$ are the curvature scale on the brane 
and the bulk anti-deSitter curvature scale, respectively.

\subsection{0th order}

It is easy to obtain the zeroth order solutions. Without derivation we present
them.
%
\begin{eqnarray}
\stac{(0)}{K^\mu_\nu}=-\frac{1}{\ell}\delta^\mu_\nu,
\end{eqnarray}
%
%
\begin{eqnarray}
\stac{(0)}{g}_{\mu\nu}=a^2(y,x)h_{\mu\nu}(x)=e^{-\frac{2d(y,x)}{\ell}}h_{\mu\nu}(x),
\end{eqnarray}
%
where 
%
\begin{eqnarray}
d(y,x) = \int^y_0 dy e^{\phi(x)},
\end{eqnarray}
%
%
\begin{eqnarray}
\frac{1}{\ell}=-\frac{1}{6}\kappa^2 \gamma_{(+)}=\frac{1}{6}\kappa^2 \gamma_{(-)}
:=-\frac{1}{6}\kappa^2 \gamma,
\label{tension}
\end{eqnarray}
%
and
%
\begin{eqnarray}
2\Lambda +\frac{5\kappa^4}{6}\gamma^2=0. \label{tune}
\end{eqnarray}
%
$\ell$ is the curvature scale of anti-deSitter like spacetimes. 
Eqs. (\ref{tension}) and (\ref{tune}) represent the Randall-Sundrum tuning 
and then the tension $\gamma_{(+)}$ and $\gamma_{(-)}$ have the same magnitude with 
opposite signature, $\gamma_{(+)}=-\gamma_{(-)}<0$. In this tuning, 
the brane geometry could be four dimensional Minkowski spacetime. 

In addition,
%
\begin{eqnarray}
\tilde G_{y\alpha_1 \alpha_2 \alpha_3 \alpha_4}=-a^4 \kappa^2 
\gamma e^\phi \epsilon_{\alpha_1 \alpha_2 \alpha_3 \alpha_4},
\end{eqnarray}
%
where $\epsilon_{\alpha_1 \alpha_2 \alpha_3 \alpha_4}$ is the Levi-Civita 
tensor with respect to the induced metric $h_{\mu\nu}$ on the 
brane.

\subsection{1st order}

Since $\chi = O({\cal F}^2)$ and the contribution of $\chi$ to gravitational field equation is 
appeared 
as form $(\nabla \chi)^2 = O({\cal F}^4)$, we will omit such terms which will be important in the next 
order. 

The first order equations for $\tilde F_{y\mu\nu}$ and $H_{y\mu\nu}$ are 
%
\begin{eqnarray}
\partial_y \stac{(1)}{\tilde F}_{y\mu\nu}-\frac{1}{2a^4}
\stac{(1)}{H}_{y\alpha\beta} \tilde G_{y\rho\sigma\mu\nu}h^{\alpha\rho}h^{\beta\sigma}=0,
\end{eqnarray}
%
and
%
\begin{eqnarray}
\partial_y \stac{(1)}{H}_{y\mu\nu}+\frac{1}{2a^4}\stac{(1)}{\tilde F}_{y\alpha\beta}\tilde 
G_{y\rho\sigma\mu\nu}h^{\alpha\rho}h^{\beta\sigma}=0.
\end{eqnarray}
%
Together with the junction conditions on $D_+$-brane the solutions are given by 
%
\begin{eqnarray}
\stac{(1)}{H}_{y\mu\nu}(y,x)=-\kappa^2 \gamma a^{-6}e^\phi {\cal F}_{\mu\nu}^{(+)},
\end{eqnarray}
%
and
%
\begin{eqnarray}
\stac{(1)}{\tilde F}_{y\mu\nu}(y,x)=-\frac{\kappa^2}{2}\gamma a^{-6} e^\phi 
\epsilon_{\mu\nu\rho\sigma}{\cal F}_{\alpha\beta}^{(+)}h^{\rho\alpha}h^{\sigma\beta}.
\end{eqnarray}
%
The remaining junction conditions on $D_-$-brane imply the relation between ${\cal F}^{(+)}_{\mu\nu}$ 
and ${\cal F}^{(-)}_{\mu\nu}$ as 
%
\begin{eqnarray}
{\cal F}_{\mu\nu}^{(-)}=a_0^{-6}{\cal F}_{\mu\nu}^{(+)},
\end{eqnarray}
%
and then 
%
\begin{eqnarray}
T_{\mu\nu}^{(-)}= a_0^{-14} T_{\mu\nu}^{(+)} \label{emtensor},
\end{eqnarray}
%
where $a_0=a(y_0,x)=e^{-d_0(x)/\ell}$ and $d_0(x):=d(y_0,x)$. 

Let us first 
substitute the junction conditions for $H_{y\mu\nu}$ and $\tilde F_{y\mu\nu}$ on the $D_+$ brane 
into the constraint equations of Eqs. (\ref{con1}) and (\ref{con2}). Then we see 
%
\begin{eqnarray}
& & {\cal D}^\mu {\cal F}_{\mu\nu}^{(+)}=0, \label{max1} \\
& & \epsilon^{\mu\nu\alpha\beta} {\cal D}_\nu 
{\cal F}^{(+)}_{\alpha\beta}=0, \label{max2}
\end{eqnarray}
%
where ${\cal D}_\mu$ is the covariant derivative with respect to $h_{\mu\nu}$. 

Here we remind that the consistency of the assumptions of $H_{\mu\nu\alpha}=\tilde F_{\mu\nu\alpha}=0$ 
implies $\phi=$constant. To see this we begin with the identity 
%
\begin{eqnarray}
\epsilon^{\beta\alpha\mu\nu}D_\alpha \tilde F_{y\mu\nu} =0,
\end{eqnarray}
%
which is easily derived from $\tilde F_{\mu\nu\alpha}=0$ and the definition of $\tilde F_{MNK}$. Then 
we put the junction conditions on the $D_+$-brane into the above, that is, 
%
\begin{eqnarray}
\epsilon^{\beta\alpha\mu\nu}{\cal D}_\alpha(e^\phi \epsilon_{\mu\nu\rho\sigma}{\cal F}^{(+)\rho\sigma}) =0.
\end{eqnarray}
%
Using Eq. (\ref{max1}) we can see that the above equation becomes 
%
\begin{eqnarray}
{\cal F}^{(+)\alpha\beta}{\cal D}_\alpha e^\phi =0.
\end{eqnarray}
%
Therefore $\phi=$const. is required under the assumption of $H_{\mu\nu\alpha}=\tilde F_{\mu\nu\alpha}=0$. 
From now on we will set $\phi=0$ without loss of generality. 

%
%
%
%

Using these results the evolutional equation for the traceless part 
of the extrinsic curvature is 
%
\begin{eqnarray}
e^{-\phi} \partial_y \stac{(1)}{\tilde K^\mu_\nu}  =  
-\stac{(0)}{K}\stac{(1)}{\tilde K^\mu_\nu}+
\frac{^{(4)}\tilde R^\mu_\nu (h)}{a^2}{} 
 -\kappa^4 \gamma^2 a^{-16} T^{(+)\mu}_{~~~\nu}, 
\end{eqnarray}
%
where ${}^{(4)}R^\mu_\nu (h)=h^{\mu\alpha}{}^{(4)}R_{\alpha\nu}(h)$ 
is the Ricci tensor with respect to $h_{\mu\nu}$
and $T^\mu_\nu= h^{\mu\alpha}T_{\alpha\nu}$. 

The solution is summarised as 
%
\begin{eqnarray}
\stac{(1)}{\tilde K^\mu_\nu}(y,x)  =  
-\frac{\ell {}^{(4)}\tilde R^\mu_\nu (h)}{2a^2} 
+\frac{\kappa^2 \gamma}{2}  a^{-16}T^{(+)\mu}_{~~~\nu} 
+ \frac{\chi^\mu_\nu (x)}{a^4}, 
\label{sol}
\end{eqnarray}
%
where $\chi^\mu_\nu$ is the ``integration of constant". 
The solution to the trace part of the extrinsic curvature is 
%
\begin{eqnarray}
\stac{(1)}{K}(y,x) & = & -\frac{\ell}{6a^2} {}^{(4)}R(h). \label{trace}
\end{eqnarray}
%

Using the junction conditions on the $D_+$-brane Eqs (\ref{sol}) and (\ref{trace}) becomes 
%
\begin{eqnarray}
{}^{(4)}\tilde R^\mu_\nu (h) = \frac{2}{\ell} \chi^\mu_\nu (x) ,
\end{eqnarray}
%
and
%
\begin{eqnarray}
0=\stac{(1)}{K}(0,x)=-\frac{\ell}{6} {}^{(4)}R(h).
\end{eqnarray}
%
They correspond to the Einstein equation on the brane obtained in Ref. \cite{SMS} and 
$ \chi^\mu_\nu$ is projected Weyl tensor $E_{\mu\nu}$. For the moment, 
$\chi^\mu_\nu (x)$ is unknown term. 

On $D_- $-brane, Eq. (\ref{sol}) becomes 
%
\begin{eqnarray}
\frac{\kappa^2 \gamma}{2} T^{(-) \mu}_{~~~\nu} =  
-\frac{\ell {}^{(4)}\tilde R^\mu_\nu (h)}{2a_0^2}  
+\frac{\kappa^2 \gamma}{2}a_0^{-16} T^{(+)\mu}_{~~~\nu}+\frac{\chi^\mu_\nu (x)}{a_0^4}.
\end{eqnarray}
%
Using Eq. (\ref{emtensor}) the energy-momentum tensor in both sides are exactly canceled out and then 
%
\begin{eqnarray}
\chi^\mu_\nu (x) = \frac{\ell}{2}a_0^2 {}^{(4)}\tilde R^\mu_\nu (h). 
\end{eqnarray}
%

All together we obtain the Einstein equation on $D_+$ brane 
%
\begin{eqnarray}
(1-a_0^2){}^{(4)}G_{\mu\nu} (h) =0.
\end{eqnarray}
%
This is main result in our paper. The gauge fields do not appear in the right-hand side. However, 
there is the equation for the gauge field obtained from the constraint equations(See Eqs. (\ref{max1}) and 
(\ref{max2})).

\section{Summary and discussion}
\label{sec:summary}

We considered the two D-brane system under the Randall-Sundrum tuning. 
Therein U(1) gauge fields on the brane plays role as the source 
for the three form fields in the bulk. Then we derived the field equations on the brane after solving 
the bulk field equation with junction condition in the two brane system
where the long wave approximation is valid. Consequently 
we find that the gravity on the brane does not couple to the gauge field. The result is same with the 
single D-brane case. Under the assumptions of $H_{\alpha\beta\mu}=\tilde F_{\alpha\beta\mu}=0$, 
the physical mode, radion, which expresses the distance between two branes is 
required to be constant. From our procedure, we can expect that the same result will be 
obtained for Yang-Mills field localised on the brane. 

The result we obtained here is quite unusual. However, we do not know the situation where the effect of 
the gauge field to the gravity is important except for homogeneous and isotropic universe. 
In standard big-bang scenario the universe is dominated by the radiation before the equal time. 
The success of the big-bang scenario suggests the expansion of the universe is governed by 
unknown exotic matter like dark energy if one is serious about our result. 
On the other hand, we know the situation where 
the effect of the fermion to the gravity is important. If we see the similar result for fermions, 
it will be crisis in D-braneworld cosmology. There is a solution to this problem. 
As guessed in Ref. \cite{SKT}, the net cosmological constant on the brane might be important and its 
presence implies the coupling of the gravity to matter. 


\vskip -1cm 
\section*{Acknowledgements}

The work of TS was supported by Grant-in-Aid for Scientific
Research from Ministry of Education, Science, Sports and Culture of 
Japan(No.13135208, No.14740155 and No.14102004). 
The works of YH and KT were supported by JSPS.


\end{document}